# Induced Matter Theory & Heisenberg-like Uncertainty Relations

James R. Bogan
2925 SW 49th St.
Corvallis, OR 97330
Jim_bogan@hotmail.com

**Abstract**

We show that a differential variant of the Heisenberg uncertainty relations emerges naturally from induced matter theory, as a sum of line elements in both momentum & Minkowski spaces.

**Introduction**

Just after the introduction of Kaluza-Klein theory in the 1920's [1,2], Campbell [3] and later Magaard [4] have argued that it is possible to embed a n-dimensional theory into a n+1 dimensional manifold. Today, a few workers are using this basic notion to investigate a wide range of theoretical topics, from brane worlds[5-7] and cosmology [8-11], to relativistic quantum mechanics [12-14].
Induced matter theory posits that matter is in fact a direct manifestation of a non-compactified, fifth spatial dimension [15].

Since matter is intrinsically quantum-mechanical (QM), we argue here that the origin of QM uncertainties in energy-momentum and space-time, must ultimately originate in the fifth dimension, and are thus geometric in nature. As Moffat has recently argued [16], there exist completely dualistic descriptions of classical particle motion, such that one may construct all the mathematical elements of general relativity, e.g., line elements, affine connexions, curvature tensors, space-time tensors, etc., in momentum space, in a fashion exactly analogous to a psuedo-Riemannian space.

Here, we extend this program, and argue that the propagation of a complex scalar field in a 5-D space-time continuum, leads to both a 5-momentum conservation condition and in turn, to a geometrical origin of the Heisenberg uncertainty relations.

Index convention shall be Greek for 3+1 dimensions (e.g., $\mu = 0 - 3$) and Latin for 4 +1 dimensions (e.g., $A = 0 - 4$), with metric signatures (+---) and (+----) respectively.

**Theory**

We begin by investigating the propagation of a massive, complex scalar field, via the Klein-Gordon equation (KGE),

(1) $\partial^\mu \partial_\mu \Psi + k_c^2 \Psi = 0$, Where $k_c = \dfrac{mc}{\hbar}$ is the Compton wavenumber & $m$ is the mass of the scalar field quantum. In 5-D, the KGE goes over to the null D'Alembertian,

(2) $\partial^A \partial_A \Psi = 0$ in which the Compton wavenumber is the eigenvalue of the 5$^{th}$ dimensional term $\partial^l \partial_l \Psi = -k_c^2 \Psi$. Representing the field as a plane wave, with $\Psi = \exp(iS/\hbar)$, and identifying S as the classical action, eq.(2) then becomes,

(3) $\partial^A S \partial_A S + i\hbar \partial^A \partial_A S = 0$

Defining the 5-momenta as $P_A = \partial_A S$, we rewrite (3) as,

(4) $P^A P_A + i\hbar \partial^A P_A = 0$ Since the real & imaginary parts of this invariant must vanish individually, eq.(4) must have a null real part,

(5) $P^A P_A = 0$ Expanding the sum gives the energy-momentum relation of special relativity,

(6) $P^\mu P_\mu = P^l P_l = (mc)^2$ Similarly, the imaginary part can be expanded & written as

(7) $\partial^\mu P_\mu = \partial^l P_l = \eta$ Where $\eta = \dfrac{(mc)^2}{\hbar}$ is a constant required for dimensional consistency.

Rewriting eq. (7), and forming a bilinear null invariant,

(8) $(dP^\mu - \eta dx^\mu)(dP_\mu - \eta dx_\mu) = 0$ , Expanding & solving for the cross-term, gives

(9) $dx^\mu dP_\mu = \dfrac{1}{2\eta}(dP^\mu dP_\mu + \eta^2 dx^\mu dx_\mu)$, redefining the two terms on the RHS as dimension-less, differential invariants,

(10) $\quad d\rho^2 = \dfrac{dP^\mu dP_\mu}{(mc)^2}$ , and $\quad d\chi^2 = k_c^2 dx^\mu dx_\mu \quad$ we can write eq.(9) as

(11) $\quad dx^\mu dP_\mu = \dfrac{\hbar}{2}(d\rho^2 + d\chi^2)$

Eq.(11) is a Heisenberg uncertainty-like relation, in which the mean value of the two invariant line elements in canonically-conjugate metric spaces, contributes to a total uncertainty in four-momentum and -position, scaled by Dirac's action constant.

Conclusion

From the physics of STM theory, we have shown that the propagation of a complex scalar field in 5-D, leads directly to an invariant with a form similar to the Heisenberg uncertainty relations. Moreover, the `backbone' of these relations originates in the line elements of the differential geometries of momentum and Minkowski spaces. This suggests that there may be a common descriptor to both the geometric and probability notions of relativistic quantum mechanics.